%% file: offset.tex
\documentclass[useAMS,usenatbib,usegraphicx]{mn2e}
\newcommand{\swift}{\emph{Swift}}

\newcommand{\ee}[1]{$\times$10$^{#1}$}

\newcommand{\msun}{M$_{\odot}$}

\title[Progenitors of short hard gamma--ray bursts]
{Different progenitors of short hard gamma--ray bursts}

\author[Eleonora Troja et al.]{E.~Troja$^{1,2,3}$, A.~R.~King$^{1}$,
                         P.~T.~O'Brien$^{1}$, N. Lyons$^{1}$ and
                         G.~Cusumano$^{2}$ 
\\
$^{1}$Department of Physics and Astronomy, University of Leicester,
                 Leicester, LE1~7RH, UK \\ $^{2}$INAF - Istituto di
                 Astrofisica Spaziale e Fisica Cosmica, Sezione di
                 Palermo, via Ugo la Malfa 153, 90146 Palermo, Italy
                 \\
$^{3}$Dipartimento di Scienze Fisiche ed Astronomiche,
                 Sezione di Astronomia, Universit\`a di Palermo, 
		 Piazza del Parlamento 1, \\ 90134 Palermo, Italy}
\begin{document}

\date{Accepted -- -- --. Received -- -- --.}

\pagerange{\pageref{firstpage}--\pageref{lastpage}} \pubyear{2007}

\maketitle

\label{firstpage}

\begin{abstract}

We consider the spatial offsets of short hard gamma--ray bursts (SHBs)
from their host galaxies. We show that all SHBs with
extended--duration soft emission components lie very close to their hosts.
We suggest that NS-BH binary mergers offer a natural explanation for
the properties of this extended--duration/low offset group. 
SHBs with large offsets have no observed extended emission components 
and are less likely to have an optically detected afterglow, 
properties consistent with NS-NS binary mergers occurring in 
low density environments.

\end{abstract}


\section{Introduction}\label{sec:intro}

\input{tab1.tex}


\begin{figure*}
\centering
\includegraphics[angle=0,scale=0.49]{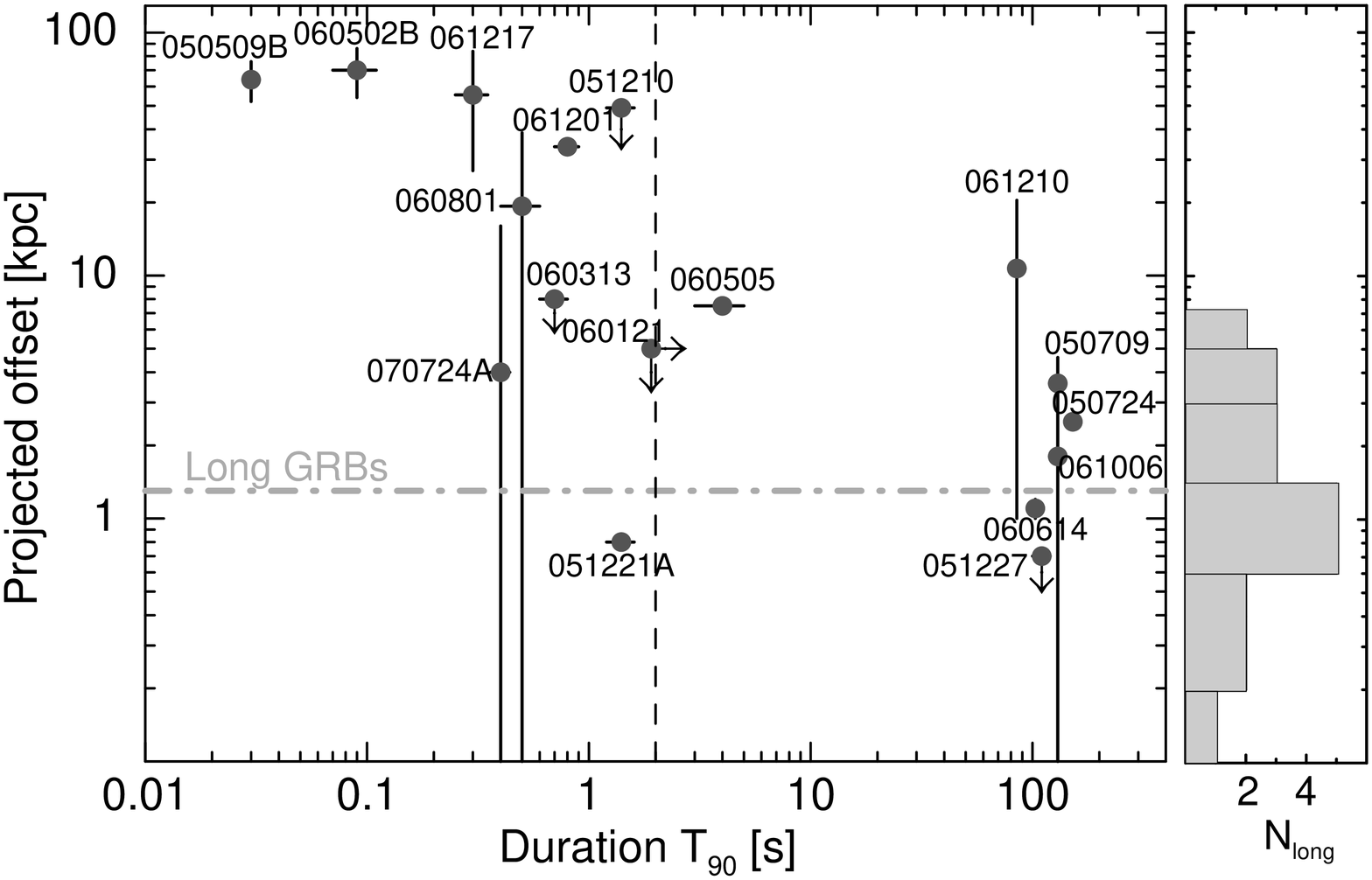}
\caption{{\it Left panel:} projected physical offsets as a function of the burst
         duration (T$_{90}$) in the $\gamma$-ray band.  The vertical
         dashed line marks the canonical temporal division between
         long and short hard bursts.  The horizontal dot-dashed line
         reports the median offset for a sample of long GRBs
         with known redshift (from Bloom et al. 2002). 
	 {\it Right panel:} Offsets histogram for the same sample of long GRBs.}
\label{f1}
\end{figure*}


In the last few years, the successful {\it Swift} mission
\citep{swift04} has greatly expanded our knowledge of gamma-ray burst
(GRB) phenomenology. In particular, it has transformed the study of
SHBs. The ability to react rapidly to GRBs triggers led to
the first detection of a SHB X-ray afterglow (GRB~050509B;
\citealt{gehrels05}), and, a few months later, to the detection of the
first SHB optical counterpart (GRB~050709;
\citealt{fox05,hjort05}). Accurately pinpointing the afterglow
position on the sky can link the SHB to its host galaxy, constraining
its distance and energetics through the galaxy's redshift
measurement. Identifying SHB hosts can also provide a powerful insight
into the progenitor population and formation history. 
Almost all SHB models invoke close binary systems containing at 
least one neutron star. The mass loss involved in the supernova 
forming the neutron star gives the binary a significant space velocity, 
depending on its total mass. This can be enhanced if the back reaction 
(`kick') on the neutron star is anisotropic. 
There is ample observational evidence 
(e.g. \citealt*{wang06} and references therein) 
for such anisotropic kicks in both single and binary neutron stars.

The analogous inferences for long GRBs \citep{bloom02,fruchter06} are
well known. For instance, only a few important cases show an observed
GRB/supernova (SN) connection (e.g. GRB~060218/SN2006aj;
\citealt{campana06b,pian06}), but the measured low offsets from the
galaxy centres and the preferential location of long GRBs in the
bluest regions of these galaxies strengthen the link with massive
stars and their collapse. By contrast, associating SHBs with a host is
complicated by the faintness of their afterglows and their potential
origin in NS binaries which can travel far from their birth sites
before coalescence (\citealt*{bloom99}; \citealt{belcz06}; 
\citealt{wang06}; \citealt*{lee07}). Finding
the absorption redshifts of SHB afterglows would strengthen the
association with their hosts.

Since its launch, in November 2004, {\it Swift} has detected 25 GRBs
classified as SHBs up to August 2007.  
In a significant fraction of them ($\sim$25\%)
the initial short hard $\gamma$-ray episode is followed by a second
spectrally softer emission component, lasting tens of seconds.
Despite their long duration, exceeding the canonical cut of 2~s
\citep{kouvel93}, these bursts display all the distinctive features of
the SHBs class: a first short-hard event with zero spectral lag
\citep{nb06}; a heterogeneous population of host galaxies, in stark
contrast to the hosts of long GRBs which are all late type
\citep{covino06,prochaska06}; very tight limits on the
presence of any accompanying SN, at odds with the standard
core-collapse origin of long GRBs \citep{woosley93}. 

In 18 cases out of 25 ($\sim$70\%) there is an X-ray counterpart, and
in 7 cases ($\sim$28\%) the optical afterglow was also detected.  
Three additional bursts with visible X-ray and optical counterparts, were
triggered by the {\it HETE-2} (GRB~050709, GRB~060121; 
\citealt{villasenor05,donaghy06}) and {\it INTEGRAL} (GRB~070707; \citealt{gcn6608}) satellites.  
A total of 21 SHBs have arcsecond or sub-arcsecond localizations, 
allowing us to infer their hosts and estimate their redshifts with some security.

In this Letter we report on the full sample of well-localized SHBs and
their possible progenitors, focussing on their spatial distribution
with respect to their putative hosts. We also estimate the prompt
$\gamma$-ray and X-ray afterglow energetics of the available sample.
The paper is organized as follows: in \S~2 we briefly describe the
adopted selection criteria and the general properties of the sample;
our results are reported in \S~3.  We discuss our findings and
their implication for SHBs progenitors in \S~4.  A summary of our
conclusions is given in \S~5.  Throughout the paper we have adopted a
standard cosmology with Hubble constant
$H_0$=71\,km\,s$^{-1}$\,Mpc$^{-1}$ and parameters
$\Omega_\Lambda=0.73$, $\Omega_M=0.27$ \citep{wmap}.

\section{Sample}

We include in our analysis GRBs whose prompt emission follows the
original classification (T$_{90}$$<$2 s, hard spectrum;
\citealt{kouvel93}), as well as GRBs that formally have a long duration
(T$_{90}$$>>$2 s), but a morphology resembling the short bursts with
extended emission, as codified by \citet{nb06}.  We discard those
GRBs without at least an accurate X-ray localization.  
Among the 21 well-localized ($\la$\,6\arcsec\ radius) SHBs,
we excluded six other bursts since their hosts and distance scales are not constrained
(GRB~050813, GRB~070429B, GRB~070707, GRB~070714B, GRB~070729, GRB~070809).

In addition two bursts, GRB~060505 and GRB~060614 \citep{fynbo06,gehrels06}, which display
several features of the SHBs class, were considered and compared to
the sample.

Table~\ref{tab:sample} lists the properties of our sample of bursts
and their putative hosts.  In each case we give the probability,
P$_{\rm chance}$, that the proposed association is a chance
coincidence (col. 5).  If no value is given in the literature, we
simply estimated it as the probability that a galaxy of magnitude
$R$$<$$R_{\rm host}$ is randomly placed within a certain radius from
the GRB centroid position, without regard to the galaxy type or
redshift.  When the galaxy centroid lies within the error circle
position (e.g. GRB~061006), then the GRB cross section is determined
by the size of the uncertainty region.  Otherwise, if the galaxy is
well outside the position circle (e.g. GRB~061217), it is determined
by the angular offset (col. 7).  We used the results of \citet{hogg97}
and \citet{huang01} to calculate the galaxy sky-density in the
$R$-band.

The derived values listed in col. 5 reflect the difficulty of
identifying SHB hosts. These result either from poor localizations or
large offsets (e.g. GRB~061217).  The chance of a spurious
association obviously increases when only an X-ray position is
available, as several galaxies lie within or close to the X-ray error
circle.  In those cases, the guiding criterion is usually the object
brightness, favouring the association with the brightest galaxy.
Interestingly, the probability that 4 associations out of 17 are
spurious is $\sim$4\ee{-4}, and indeed the chance of 4 or more
misidentifications is well below the 3\,$\sigma$ confidence level.


The quoted errors are mainly due to the GRB localizations,
usually pinpointed within a 90\% confidence level error circle.
We caution that the offset is a positive-defined quantity, thus
the associated uncertainties do not properly reflect a probability distribution, 
especially in cases of negligible offsets (see \citealt{bloom02}).

\section{Results}

Fig.~\ref{f1} presents the projected galactocentric offset of SHBs 
as a function of the burst duration in the $\gamma$-ray band (observer frame). 
For comparison, the median offset value for long bursts ($\sim$1.3~kpc;
\citealt{bloom02}) is traced by the horizontal line. The frequency
histogram of long bursts as a function of the projected offset is shown 
in the narrow right panel. 
Two main features emerge from the plot: 1) bursts with prompt emission
extending up to $\sim$100-200~s tend to be clustered very close to
their host galaxy, while short bursts display a more heterogeneous
displacement around the host; in particular 2) the shortest duration
bursts seem to prefer much higher offsets than the rest of the sample.

\begin{figure}
\centering
\vspace{0.3cm}
\includegraphics[scale=0.42]{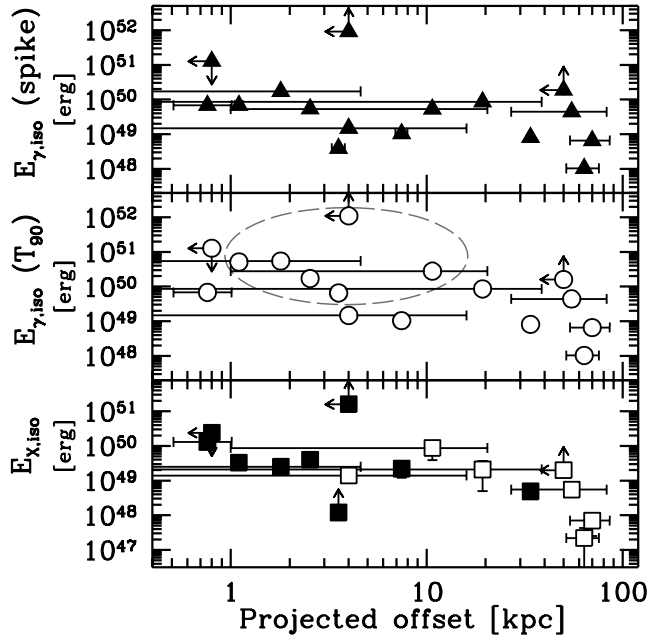}
\caption{
Prompt and afterglow energetics (source rest frame) as functions of the projected physical offset.
{\it Top panel}: Isotropic energy (15-150 keV) released 
over the initial short hard $\gamma$-ray event only (spike).  
{\it Middle panel}: : Isotropic
energy (15-150 keV) over the T$_{90}$ duration.  
Bursts with a long lasting emission are enclosed by the
dashed ellipse.  {\it Bottom panel}: Isotropic energy (0.3-10 keV)
released over the temporal range
400~s--500~ks. Filled and empty
symbols indicate GRBs with and without a detected optical afterglow,
respectively.}
\label{f2}
\end{figure}

In Fig.~\ref{f2} the prompt and the afterglow energetics are shown as
functions of offset.  In all cases we assumed isotropic emission.
The $\gamma$-ray and the X-ray energies are calculated in the 15--150
keV and the 0.3-10 keV bands respectively.  To refer our results to
the same rest frame energy band we derived a k-correction from the
burst spectral parameters (see references in Tab.~\ref{tab:sample}).

The $\gamma$-ray energy radiated during the short hard spike and over
the total T$_{90}$ are reported in the top panel and in the middle
panel of Fig.~\ref{f2}, respectively. Bursts with extended emission
are on average more energetic than bursts with T$_{90}$$<$2~s, as
shown in Fig.~\ref{f2} (middle panel), but no clear distinction
emerges if we consider only the energy of the initial hard event (top
panel).

The bottom panel of Fig.~\ref{f2} shows the X-ray isotropic energy,
calculated by integrating the best fit lightcurve between 400~s and
500~ks after the trigger (rest frame time), when the central engine
activity does not dominate the total X-ray emission.  In two cases, GRB~060801
and GRB~051210, the X-ray afterglow was below the detection limit in
this temporal range. To estimate their energetics we assumed 
temporal slope $\alpha$$\sim$1 and spectral index $\beta$$\sim$1
($F_{\nu,\rm t} \propto \nu^{-\alpha} {\rm t}^{-\beta}$).  The
normalizations were determined by the upper limits from \swift/XRT
observations.  Filled symbols indicate those bursts with a detected
optical counterpart, empty symbols those lacking an optical detection.

Even given the small number of SHBs detected so far, it is clear that
large offset bursts (GRB~050509B, GRB~060502B, GRB~061201 and
GRB~061217) lie on the lower part of the bottom panel of
Fig.~\ref{f2}, while small offset bursts instead have on average more
energetic X-ray afterglows and a much higher chance of a detectable
optical afterglow.  \citet{malesani07} noticed that optical
counterparts of SHBs with extended emission are more frequently
detected. Our Figure~\ref{f2} suggests that this is an enviromental
property, since these bursts seem to happen closer to their hosts,
and hence presumably in denser interstellar environments.

\section{Discussion}


 As shown in Fig.~\ref{f1}, short GRBs with measured offsets 
appear qualitatively divided into two groups.
The group with extended durations all lie
very close to their hosts, while the group with short duration have
a mean offset a factor of 15 larger.
Though the low statistic does not allow us
to firmly assess that the proposed groups belong to two distinct offset
distributions, a Kolmogorov-Smirnov test, ran on the current sample of bursts, 
excludes at the 2$\sigma$ confidence level
that they are drawn from the same distribution.
Furthermore, we point out that the two groups are characterized by very different 
observational features, which are hard to explain if they proginate from the same 
parent population.
 
The two groups (extended duration/small offset, short duration/large
offset) have similar redshift distributions (see Tab~\ref{tab:sample},
col. 2). Accepting the usual arguments that the short duration/large
offset group are probably NS--NS mergers, we then have four {\it a priori}
possibilities for explaining the extended duration/small offset
group. These are: a different class of NS+NS mergers, NS+massive WD
mergers, collapsars, and NS+BH mergers. We consider these in turn.

\subsection{A different class of NS+NS mergers}
The obvious possibility here is ultracompact NS--NS binaries, 
which for suitable binary kick velocities 
\mbox{$v_{\rm kick}\sim100$~km~s$^{-1}$} can produce rather
small offsets from certain types of host (cf. \citealt{belcz06},
Fig. 3). The problem here is that the initial (pre--afterglow) NS--NS
merger process should be exactly the same as for NS--NS binaries
starting from wider separations. Yet the small offset group have
rather distinct features (e.g. a prompt extended tail of emission, a higher
energetic budget) which cannot result from environmental effects.

\subsection{NS+massive WD mergers}
This group has the desirable properties \citep{king07} of
extended duration and no supernovae, but is likely to have similar
merger times and kicks as the standard NS--NS group. It therefore
cannot explain the small offsets.

\subsection{Collapsars}
Collapsars offer a simple explanation of the small offsets, but have
other problems. In particular one would have change the model (e. g. \citealt{fryer07}) to
explain both the very different light curves and the lack of supenovae
in the extended duration/small offset group. Moreover at least one
observed member of this group is hosted by an elliptical galaxy
(GRB~050724; \citealt{berger05,malesani07}), which is hard to reconcile with a collapsar origin.


\subsection{NS+BH mergers}
Low offsets are expected for NS--BH mergers on two quite general
grounds. First, there is mounting observational evidence that at least
some black holes do not receive natal kicks. \citet{mirabel03} show
that the 10 \msun\ BH binary Cyg X--1 has a peculiar velocity of
$<$10~km~s$^{-1}$, and \citet{dhawan07} show that the kick in the BH
binary GRS 1915+105 was probably similarly small. These may therefore
be examples of direct collapse to a black hole \citep{fryer01}.
(Direct collapse to a neutron star is not possible, as this has far
lower entropy than its progenitor, unlike a black hole.) Second, the
gravitational radiation merger times $t_{\rm GR}$ for BH--NS and
NS--NS binaries of a given initial separation scale as $\sim (M_{\rm
BH}/M_{\rm NS})^{-2} \sim 0.01$ for typical masses $M_{\rm BH} = 14$
\msun, $M_{\rm NS} = 1.4$ \msun.  Together these two effects show that
some BH--NS binaries would move very little before merging to produce
a short GRB.

The advantage of the latter explanation of the low offsets is of
course that it offers natural interpretations of the peculiar features
of the group of SHBs with extended emission.  \citet{rosswog07}
proposed that if a significant fraction of the shredded NS is not
immediately accreted, but remains in bound orbits around the central
object, the fallback accretion of the NS remnants can inject power up
to late times ($\la$1~d after the burst).  The derived theoretical
light curves (Fig.~3 of \citealt{rosswog07}) show that NS--BH binaries
are able to produce much higher luminosities and longer durations than
NS--NS mergers.

Figure~1 shows a further surprise, in the form of its empty 
bottom-left corner. Models of standard NS--NS mergers predict that an
appreciable fraction of such binaries are ejected far from the host,
but most remain bound to it. Thus 80\%--90\% merge within 30 kpc
according to \citealt{bloom99}. These bursts should have populated the
empty short duration/small offsets region in Fig.~1. We note that five
other very short bursts (GRB~050906, GRB~050925, GRB~051105A,
GRB~070209, GRB~070810B, T$_{90}$\,$\la$\,0.1~s) 
lack an X-ray
counterpart, despite very prompt \swift/XRT follow-up observations
(79~s, 92~s, 68~s, 78~s, and 62~s after the bursts, respectively).  We
speculate that the expected low density of the intergalactic
environment may explain the faint X-ray afterglows, placing these
X-ray dark bursts in the upper-left side of Fig.~\ref{f1}.
However, other mechanisms, related to the microphysics of the shocks
and the initial Lorentz factor, could suppress the early
X-ray emission (see \citealt{nakar07}). Also, a magnetar origin, as
debated for GRB~050906 \citep{levan07} and GRB~050925, might explain
the lack of detection.



\section{Conclusions}

The offset distribution of SHBs displays several interesting features
suggesting two types of progenitor. Most strikingly we found that
SHBs with extended soft emission (T$_{90}$$\sim$100~s) tend to remain
close to their host galaxies.
NS--BH mergers naturally account for 
these properties, although other explanations are still possible.
SHBs with large offsets have properties consistent with NS--NS mergers 
occurring in low density environments.


\section*{Acknowledgments}

This work is supported at the University of Leicester by the Science
and Technology Facilities Council, and at INAF by funding from ASI on
grant number I/R/039/04 and by COFIN MIUR grant prot. number
2005025417.  
ET acknowledges the support of the Marie Curie Spartan exchange program
at the University of Leicester. 
ARK acknowledges a Royal--Society--Wolfson Research Merit Award.
NL acknowledges support from an STFC studentship.


\bibliography{ref}

\end{document}

%% file: tab1.tex

\begin{table*}
\label{tab:sample}
\centering
\begin{minipage}{170mm}
\caption{SHBs sample properties.}
\begin{tabular}{ccccccccccc}
\hline
                &        & \multicolumn{3}{c}{Putative host}  &           &           Angular     &          &   Projected               &       &       \\ 
            {\ \ \ \ GRB\ \ \ \ \ \ }& T$_{90}$ & $z$  & $R$   & P$_{\rm chance}$ & Afterglow & offset & Error    &  offset & Error & Refs. \\            
	        & (s)      &      & (mag) &                   &           & (arcsec)       & (arcsec) & (kpc)            & (kpc) &              \\
	    (1) & (2)      &  (3) & (4)   & (5)               & (6)       & (7)            & (8)      & (9)              & (10)  & (11)         \\
\hline
050509B\dotfill  &  0.03 [0.01]  &  0.225   & 16.8 &    5.0\ee{-3}	&  X\, --\, --  &  17.87  &  3.40   &  64  	&  12	&   1--3	        \\
050709\dotfill   &   130 [7]\,\footnote{{\it Hete-2} trigger. The duration is given in the 2-25 keV energy band.}
                                 &  0.161   & 21.2 &    2.0\ee{-3}	&  X,O\, --   	&  1.30   &  0.10   &  3.57   	&  0.27	&   4--6	        \\
050724\dotfill   &   152 [9]	 &  0.258   & 19.8 &    1.0\ee{-5}	&  X,O,R   	&  0.64   &  0.02   &  2.57  	&  0.08	&   7--9        \\
051210\dotfill   &   1.4 [0.2]	 &  $>$1.4  & 23.8 &    1.0\ee{-1}	&  X\, --\, --  &  2.80   &  2.90   &  $<$50   	&  -- 	&   3, 10, 11   \\
051221A\dotfill  &   1.27 [0.05]   &  0.546   & 22.0 &    2.4\ee{-4}	&  X,O,R   	&  0.12   &  0.04   &  0.76   	&  0.25	&   12, 13	\\
051227\dotfill   &   110 [10]
				 &  --      & 25.6 &    2.0\ee{-4}	&  X,O\, --   	&  0.05   &  0.02   &  $<$0.7   &  --   &   11, 14           \\
060121\dotfill   &  1.97 [0.06]\,\footnote{{\it Hete-2} trigger. The duration is given in the 30-400 keV energy band. 
                  \citet{donaghy06} detected a faint and long-lasting soft bump of emission at a significance level of $\sim$4.5$\sigma$.}	   
 
                                 &  $>$1.7  & 26.6 &    1.3\ee{-2}	&  X,O\, --   	&  0.32
				                                                                   &  0.10   & $<$4  	&  --   &   15--17   \\
060313\dotfill   &   0.7 [0.1]   &  $<$1.1  & 25.0 &    4.0\ee{-3}	&  X,O\, --   	&  0.40   &  0.56   &  $<$8     &  --   &   11, 18	\\
060502B\dotfill  &  0.09 [0.02]  &  0.287\,?\,\footnote{A faint ($R$=26 mag) object (S2 in \citealt{bloom07}) has been proposed as the high
redshift host galaxy. The measured angular offset is 4.2$\pm$3.7 arcsec (P$_{\rm chance}$$\sim$70\%), corresponding to 34$\pm$30\,kpc at $z\sim1$.}
                                            & 18.7 &    $<$5.0\ee{-2}	&  X\, --\, --  &  16.33  &  3.70   &  70   	&  16   &   3, 19   \\
060505\dotfill   &     4 [1]	 &  0.089   & 17.9 &    1.0\ee{-4}	&  X,O\, --   	&  4.53   &  0.32   &  7.45   	&  0.53 &   20, 21	\\
060614\dotfill   &   103 [5]	 &  0.125   & 22.5 &    6.0\ee{-6}	&  X,O\, --  	&  0.50   &  --     &  1.10   	&  --   &   22--24	\\
060801\dotfill   &   0.5 [0.1]   &  1.131   & 23.0 &    4.1\ee{-2}	&  X\, --\, --  &  2.39   &  2.40   &  19.7   	&  19.8 &   3, 11	\\
061006\dotfill   &   130 [10]	 &  0.438   & 23.7 &    1.8\ee{-3}	&  X,O\, --   	&  0.32   &  0.50   &  1.8   	&  2.8  &   11, 26	\\
061201\dotfill   &   0.8 [0.1]   &  0.111   & 19.0 &    3.8\ee{-2}	&  X,O\, --   	&  17.00  &  0.20   &  33.9   	&  0.4  &   27	\\
061210\dotfill   &    85 [5]	 &  0.410   & 21.1 &    4.7\ee{-3}	&  X\, --\, --  &  1.99   &  1.80   &  10.7   	&  9.7  &   3, 11, 28	\\
061217\dotfill   &  0.30 [0.05]  &  0.827   & 23.4 &    3.9\ee{-1}	&  X\, --\, --  &  7.41   &  3.80   &  55   	&  28   &   3, 11, 29	\\
070724A\dotfill   &  0.40 [0.04]  &  0.457  & $\sim$21\,\footnote{We assume $R-I\sim1$}
                                                   &    $\sim$5\ee{-3}	&  X\, --\, --  &  0.72   &  2.10   &  4  	&  12   &   3, 30	\\

\hline 
\end{tabular}	  \\

{\sc Notes :} Col. (1): GRB name; Col. (2): T$_{90}$ duration and its error in the 15-350~keV energy band;
Col.~(3): Redshift of the putative host galaxy; Col.~(4): Observed $R$ magnitude of the putative host galaxy; 
Col. (5): Probability that the association is a chance of coincidence; Col. (6): Detection of the GRB counterpart in 
different energy band (X - X-ray; O - optical; R - radio); 
Col. (7)-(8): Angular offset between the afterglow position and the associated galaxy centroid, and its
error, respectively;
Col. (9) and (10): Projected physical offset and its error, respectively;
Col. (11): Reference to publications of the presented data.\\

{\sc Refs.:} (1)~\citealt{gehrels05}; (2)~\citealt{bloom06}; (3)~\citealt{butler07}
(4)~\citealt{hjort05}; (5)~\citealt{fox05}; (6)~\citealt{villasenor05};
(7)~\citealt{campana06}; (8)~\citealt{berger05}; (9)~\citealt{prochaska06}; 
(10)~\citealt{laparola06}; (11)~\citealt{berger07};  (12)~\citealt{burrows06}; (13)~\citealt{soderberg06};
(14)~\citealt{taka07}; (15)~\citealt{donaghy06}; (16)~\citealt{postigo06}; (17)~\citealt{levan06}; (18)~\citealt{roming06}; 
(19)~\citealt{bloom07}; (20)~\citealt{ofek07}; (21)~\citealt{levesque07}; 
(22)~\citealt{galyam06}; (23)~\citealt{gehrels06}; (24)~\citealt{mangano07}; 
(25)~\citealt{gcn5381}; (26)~\citealt{gcn5718};
(27)~\citealt{gcnr061201}; (28)~\citealt{gcnr061210}; (29)~\citealt{gcnr061217};
(30)~\citealt{gcnr070724}.
\end{minipage}
\end{table*}
